Title: Integration of TNF induced apoptosis model and quantitative proteomics data of HeLa cells - our first experience

Authors: Csilla Uličná and Jozef Uličný*

Institution: Department of Biophysics, Institute of Physics, Faculty of Sciences, P.J.Šafárik University Košice, Slovakia

Addresses and e-mails: Department of Biophysics, Jesenná 5, PF UPJŠ Kosice, 041 54 Košice, Slovakia, ulicna@pulib.sk, jozef.ulicny@upjs.sk

*corresponding author: Jozef Uličný, Department of Biophysics, University of P.J.Šafárik, Jesenná 5, SK 041 54 Košice, Slovakia, e-mail: jozef.ulicny@upjs.sk

Short title: TNF apoptosis model of HeLa cell

# Integration of TNF induced apoptosis model and quantitative proteomics data of HeLa cells - our first experience

# Abstract

The TNF initiated processes are simulated using our integrative computational model with new quantitative proteomics data obtained on HeLa cell line. In spite of fact, that the model development is based on limited experimental information, with missing data estimated by simulations and parametrization using indirect experiments on different cell lines, the actual model with HeLa experimental data provides pleasant agreement between the computationally predicted behavior and experiments. Since no significant departures from experimentally determined functionality of HeLa cells has been observed, this can be taken as encouraging sign, that substantial features of the cell line behavior are reproduced correctly and indicate new possibilities for cell-line specific computational experiments.

Keywords: HeLa cells, quantitative proteomics, TNF apoptotic pathway, systems biology model

# Introduction

Reaction kinetic models of systems biology form important tools in our effort to understand complex biological processes at molecular level. In recent decade, hundreds of such models were reported, covering growing parts of various biological processes and their regulation. Being integrative, these models are seldom created from scratch, they are gradually build and refined to cover widening range of molecules, interactions and phenomena, accommodating new knowledge as it arrives. Formally exact and quantitative, such models comprise big part of our knowledge about processes, specified in terms of time course of populations of individual molecular species. The attractiveness of this approach stems from the mathematical rigorosity, allowing formulation of very compact, computer readable and processable, internally coherent framework. Quantitative models created within the framework allow numerical simulations of the modeled system behavior, even for processes and phenomena of considerable complexity. There is now established common model description language SBML (Hucka et al., 2003), as well as the repositories of models (Li et al., 2010), (Lloyd et al., 2004) facilitating the developments, sharing and use of such models. Due to the quantitative character of the reaction kinetic models, their construction and utilization requires considerable amount of quantitative information. Since the availability of experimental data is limited and available data are sometimes contradictory, the generality of the model and its plausibility in different context is not automatically guaranteed. Even for well researched topics, with thousands of experimental papers published each year, such as apoptosis, available experimental data cannot provide full, unambiguous and direct information, required by such models. In practice, many compromises are to be made in such model constructions, starting from the model reaction and interactor setup, through parametrization of reaction rates and concentrations. Lack of generally available reliable quantitative data makes model construction slow and iterative work, where new experimental data or the application of the model on new subject (organism or cell line) generally requires adjustment and often re-parametrization of the model.

Apoptosis - most known form of programmed cell death - is studied for more than century. The errors in apoptosis regulation are amongst the principal causes of tumoral and autoimmune diseases. The understanding of the regulation of apoptotic signaling pathways and the discovery of treatment-sensitive mechanisms is seen by many as the way toward the design of more effective and target-selective therapeutic strategies (Spencer and Sorger, 2011).

Apoptosis may be induced due to changes in internal cell state or due to external stimuli originating from the cell exterior. Intrinsic stimuli capable of initiating apoptosis include stress, UV radiation, ionizing radiation, activation of oncogenes, exposition to toxins etc. All these stimuli cause permanent cellular damage or significant deterioration of cell functioning, recognized inside the cell and provoking cell death response. Extrinsic apoptosis is triggered by extracellular death ligands bound to specific trans-membrane cell death receptors of the Tumor Necrosis Factor Receptor (TNFR) family (Gonzalvez and Ashkenazi, 2010) and activating signaling pathways dependent on caspases. Both intrinsic and extrinsic pathways to apoptosis share significant amount of common components and regulatory mechanisms. Due to extensive cross-talks, the comprehensive model of extrinsic apoptotic processes must thus incorporate also significant portions of the intrinsic pathways.

The regulation of apoptosis is crucial for the correct functionality of multicellular organisms, and as such must be tightly controlled. The general mechanism of apoptosis regulation must be adaptable, to provide cell-line and cell-state specific responses for the same signal. Obviously, this requires robust, reliable and tunable mechanism, where quantitative differences on input translate into qualitatively different outcomes. Apparently, in order to understand the behavior of apoptotic machinery, the qualitative description of the mechanism should be accompanied by the quantitative treatment. Due to complexity of such models, predictions usually cannot be made analytically, relying instead on numerical simulations.

It is thus of significant importance to be able to test the usability of the quantitative models in situations, where new experimental data appear, unknown at the time of model construction.

Very recently, quantitative proteomics data were published for human U2OS (Beck et al., 2011) and HeLa cell lines (Nagaraj et al., 2011). This independent experimental set can be used for the critical evaluation of the functionality of computational models, created without this information. The computational models of systems biology work with absolute quantities of the molecules and their time evolution. This information is not readily available, since most experimental data are relative, reporting only proportions between the species and the measurements are indirect, relying on experimental estimations of intermediates and proxies, with interpretation based on model assumptions. As a result, experimental data used in computational model are vulnerable to sample preparation and interpretation artifacts. Experimental protocols allow usually to trace only few of the parameters of interest, while the integrative models work with much bigger sets. The quantitative proteome represents the most direct estimations of the quantities appearing in models - the counts of interacting molecules in their reaction volumes. The appearance of quantitative proteomics measurements for the complex human cell lines allows, for the first time, to address systematically the question of model sufficiency and its predictive power. Comprehensive quantitative experimental data obtained on the single cell line under consistent experimental conditions provide opportunity to take a fresh look at known predictive models of apoptotic processes initiated through TNF (Koh and Lee, 2011), (Schliemann et al., 2011) when integrated into a single model (Fig. 1.).

The combination of new comprehensive data and their use together with our quantitative model provides interesting test of the quality and applicability of the model and the positive results provide solid ground for further modeling work as they encourage to explore other experimental sources of less reliable nature (such as micro-array data) and extrapolate those data in new, adapted models with higher confidence.

# Material and methods

The construction of the quantitative model is formidable task, seldom created from scratch. There is clear evolution of the models, where the subsequent models often build on previous work of other authors, taking substantial part of the reaction networks and sharing collected quantitative data, such as reaction rates and concentrations. The construction of our model is no exception. Recently published model of TNF signalization by Schliemann et al. (Schliemann et al., 2011) provided core interaction patterns to our integrative model of apoptosis, describing the death receptors part. The interactions of the model of Schliemann et al. described the formation of signaling complex I and II, the internalization of the complexes into cytosol, proliferatory and apoptotic blocks as well as comparatively detailed activation of caspase–8 and p43-FLIP fragment. Their study was based specifically on KYM1 cells, so the application of the model to more general treatment is not automatic. The sensitivity of KYM1 cell line (derived from human rhabdomyosarcoma) to TNF stimulus is not affected by excessive production of Bcl–2 or Bcl-xL proteins (Grell et al., 1999). KYM1 are cells of type I, i.e. the activated apoptosis pathways are not dependent on mitochondria, and as such, the intrinsic pathway part of the model is missing in this model. KYM1 cells are highly responsive to TNF stimuli, even in the absence of metabolic inhibitors, including the inhibitors of DNA synthesis (Grell et al., 1994), allowing for rapid development of apoptosis.

In order to provide more comprehensive model of TNF initiated apoptosis, we have made several modifications and extensions, documented below. The model of Schliemann et al. was complemented by mitochondria-specific part, where the structure of interactions and parametrization follows closely the seminal paper of Albeck et al. (Albeck et al., 2008b) recently actuated by explicit models of protein synthesis and degradation as described by Spencer (Spencer et al., 2009) and Aldridge (Aldridge et al., 2011).

# Model construction

The resulting mathematical model contains 84 molecules (including the intermediate forms) and 156 reactions, including those describing the synthesis of mRNA and translation of selected proteins, transport processes between the mitochondria, cytoplasm and nuclei, as well as *de novo* synthesis and degradation of proteins. The schematic of the model, comprising most important functional blocs, is depicted on Fig.1. For clarity, some intermediates, protein synthesis and degradation reactions are not plotted. All reactions are of mass action kinetics type. The molecular interaction network, in terms of nodes and topology is specified in Table 1.

The initial concentrations of species used in the model are taken from the experimental proteomics data (Nagaraj et al., 2011). The use of this data is based on assumption that the HeLa cells are in the survival steady state, i.e. without executing the apoptotic program. Mitochondria are intact and the molecules are not modified by excessive phosphorylation or cleaved. This is reflected by the numerical values for phoshorylated, cleaved molecules and majority of intermediate products set to zero.

The mass spectrometry (MS) data contain no information about the FLIP and BAR molecules. Since FLIP and BAR play important role in previous models of apoptosis (Albeck et al., 2008b) (Spencer et al., 2009), this point requires some discussion. Here we expect the low concentrations, too low to be detected reliably by MS and reported in the data set. This corroborates the assumption that the cell culture was in the pro-survival state. The stimulation by TNF is known to cause rapid synthesis of FLIP due to NFkB activation. Another critical component, the TNFR1 receptor, is also missing in the MS data. For TNFR1 this points to known deficiency of MS for membrane-bound proteins, as commented in the original experimental MS paper. It is generally known (Jupp et al., 2001) that HeLa cells contain about 3000 TNFR1 receptors on cell surface. The production rates of newly synthesized TNFR1 in cytosol was set to balance the rate of externalization and degradation of the externalized receptor, adopted from Schliemann et al., in order to keep the number of receptors for HeLa external membrane at this value.

## Compartment size estimation

Since source models specify amounts of species in concentrations, rather than the preferable absolute counts of molecules (numerical concentrations), the integration of simulation data from other models into our model requires to make assumptions about the reaction volumes. The reported size of HeLa cells is between 1200 to 4290 $\mu m^3$ (Zhao et al., 2008). For our simulations, we have used an average volume of 2425 $\mu m^3$. The volume of nucleus 375 (Maul and Deaven, 1977) to 690 $\mu m^3$ (Monier et al., 2000) corresponds to 15–28 % of the total cell volume. In our simulations, we have opted for the reference volume of nucleus at the 25% of the total cell volume. The typical volume of mammal cell mitochondria is 0.5 $\mu m^3$ (Mitochondria volume - Mammalian tissue culture cell - BNID 100438). The number of mitochondria in the cell varies, depending on the actual phase of cell cycle. In early phase of G1 cycle, is the number of mitochondria in HeLa about 400, later, in S and G2 phase it rises to 500–600 (Posakony et al., 1977). Only 5% of the HeLa population is in mitosis, 95% are thus in the interphase (G1+G2+S). Since in the interphase the cell in the state G1 spends roughly the same time as in the states G2+S, we have utilized the average number of mitochondria 500. This corresponds to the volume of mitochondria in the range 200–300 $\mu m^3$ (8–12% of the total cell volume). In further simulations, we have set the reference volume of mitochondria to 10% of the total cell volume.

## Estimation of reaction rates

Whenever available, kinetic reaction parameters were transferred from the existing partial models (Schliemann et al., 2011), (Albeck et al., 2008b), (Spencer et al., 2009), (Aldridge et al., 2011) and converted to the same units (suitable for simulations using numerical concentrations) and for reference volumes. Resulting reaction rates, obtained by reaction speeds for individual compartments were normalized to the compartment volumes (the volume of nuclei and the total volume of mitochondria) relative to the total volume of cytosol, according to methodology of Albeck (Albeck et al., 2008b). Thus obtained reaction rates are summarized in the Tables 2 and Table 3.

## Initial concentrations

Protein concentrations were taken unmodified, from the comprehensive quantitative proteomics data published recently by Nagaraj et al. (Nagaraj et al., 2011) for the HeLa cells. The size of the stimulus and the actual amount of mRNA were considered as independent variables driving the model. The mRNA amounts for selected proteins were used those of Schliemann et al. (Schliemann et al., 2011) without modification. Although the data are obtained for KYM1 cell line, numerical simulations showed, that the initial amount of mRNA varied over one order of magnitude up or down had only marginal effect on the onset of apoptosis. The quantitative proteomics data provide only the total number of NFkB and IkBa molecules, the information of the partitioning of these molecules into complexes and intermediates must be determined by other means. The values for NFkB, NFkB_N, IkBa, IkBa_N, IkBa:NFkB and IκBα:NFkB_N complexes were set to respect the experimental finding, that 9% of the total NFkB and 17% of the IkBa is present at nuclei without TNF stimulation (Birbach et al., 2002) and taking into account that proteomics data report concentration of monomeric units, while the NFkB used in simulations is dimer. The same considerations were applied for the IKK protein, being a trimer. When protein is composed from several components reported as standalone units in MS data, the amount was estimated according to the most deficient compound. E.g. for IKK, composed of IKK1 (3500 molecules per cell (#/cell)), IKK2 (6800 #/cell) and NEMO (34000 #/cell), the amount of IKK was set to 3500 copies per cell. The numerical concentrations are listed in Table 3.

The component models and functional blocks were numerically simulated and partial model results compared with the original simulation data to detect eventual misprints and omissions. All simulations were performed using Copasi (Hoops et al., 2006) framework and continuous deterministic time propagation algorithm (LSODA). During the simulations no critical volumes and concentrations were encountered requiring the more rigorous, though time consuming stochastic kinetics approach. The complete model is available in the Copasi format, which forms a superset of the SBML standard.

# Results

Thus constructed model provides complex simulation environment capable of providing numerous computational experiments. In the result section we will describe few of the relevant characteristics of the model fed with experimental HeLa cells parameters, together with the references to the corresponding experimental behavior. Most attention will be payed to the sequence of events initiated after the stationary cell is exposed to proapoptotic signal TNF ligand. In the present paper, we focused on permanent stimulation, i.e. the situation, where the fixed number of TNF ligand is placed into cell exterior and then this number is decreased only due to internalization of the receptor complex.

# The creation of signaling complex I

TNF binding to the receptor TNFR1 leads to rapid formation of the transmembrane signaling complex I. Allosteric changes on the cytosolic side of receptor, induced by ligand, attract adaptor proteins TRADD, RIP1 and ubiquitin E3 ligase TRAF2. The presence of RIP1 protein is essential for the further propagation of the signal proceeding through binding and activation of IKK - following the proinflamatory and antiapoptotic NFkB signaling pathway. Activated IKK phosphorylates IkBa proteins in the complex IkBa:NFkB. Phosphorylated IkBa undergoes ubiquitination and subsequent degradation (by proteasome, in model only implicit), releasing thus NFkB. NFkB is then translocated into a nucleus, where it acts as a transcription factor, initiating the expression of many proinflamatory, antiapoptotic (XIAP, FLIP) and regulatory genes (represented in the model by A20).

In actual HeLa simulations, the experimentally determined initial concentrations of adaptor proteins RIP1, TRADD, TRAF2 are low, compared to generic model assumptions (Table 4 ). This limits the speed of formation of both signaling complexes I and II and provides the upper limit for their absolute numbers. Compared to the state, when the cell contains sufficient number of adaptor proteins, the activation of NFkB pathway is delayed in time and of lower intensity (see Fig. 4. A, B). The formation of the signaling complex II, which activates the caspase–8 is slow. NFkB-independent synthesis of XIAP and FLIP, in later stages enhanced through NFkB activation, cause higher levels of XIAP and FLIP compared to the case when sufficient numbers of adaptor proteins are present. Thus, when adaptor protein levels are low, the activation of caspase–8 is delayed.

# Internalization of the signaling complex I and the creation of signaling complex II

The complex I is internalized (by clathrin-mediated endocytosis - details not modeled explicitly) within few minutes after the ligand binds TNFR1 receptor. NFkB signalization ceases to exist after ubiquitination and degradation of RIP1 and removal of TRAF2 from the complex I due to activity of ubiquitin-E3 ligases (Heyninck et al., 1999), (Liao et al., 2008), (Zhao et al., 2008), (Chen, 2012). Signaling complex II is formed by mutual entrapment of FADD, TRADD, FLIP and procaspases–8 and –10, forming thus DISC complex associated with the receptor (Micheau and Tschopp, 2003). The formation of DISC complex is detected already 3 minutes after the TNF stimulation and the complex remains associated with the receptor even after 60 minutes (Schütze and Schneider-Brachert, 2009).

# Activation of initiation caspases–8 and –10

Instead of procaspases–8 and –10, catalytically inactive procaspase–8 homologue FLIP can bind to DISC. FLIP is has higher binding affinity than procaspase–8, though procaspase–8 is more abundant in cells (Neumann et al., 2010). Procaspase–8 interacts with procaspase–10 and FLIP, with FLIP being preferential interaction partner (Irmler et al., 1997), (Krueger et al., 2001), (Wang et al., 2001). FLIP is processed by caspase–8 to provide p43-FLIP, which binds and activates IkB kinase (IKK) (Lavrik et al., 2003). Activated IKK is known to phosphorylate and thus inhibit IkBa. IkBa downregulates NFkB, so the activation of IKK results in induction of NFkB-mediated transcription (Scheidereit, 2006). Computer simulations and experiments indicate, that both signaling pathways (proapoptotic - caspase–8 dependent, as well as cell-survival - NFkB dependent) are activated concurrently and the final outcome is determined by the fine balance between the actual cellular levels of FLIP and initiator caspases. Neumann et al. (Neumann et al., 2010) showed the FLIP can switch off completely, induce or suppress the activation of NFkB, depending on whether the FLIP level is low, intermediate or high. The counterintuitive inhibition of NFkB for high levels of FLIP is due to inhibitory effect of high FLIP concentrations on production of p43-FLIP (Neumann et al., 2010).

It has been documented that excessive concentration of FLIP limits the access of procaspase–10 to the procaspase–8 on DISC complex (Micheau and Tschopp, 2003). For cells with lower levels of FLIP, the lower affinity procaspase–10 associates with procaspase–8. This indicates, that the capturing of FLIP and procaspase–10 on signaling complex II are mutually exclusive, in other words they do not interact together. Procaspase–10 interacts only with procaspase–8 and itself, which points to completely different mechanism of activation compared to the caspase–8 (Micheau and Tschopp, 2003), (Feoktistova et al., 2011).

## Apoptosis type I or type II

For extrinsic apoptotic pathway, two distinct mechanisms are described - type I and type II. Type II or automacrophagic cell death is attributed to cells with profound defects in apoptotic cell machinery (Tsujimoto and Shimizu, 2005). While more conventional apoptosis of type I doesn't require participation of mitochondria, in apoptosis type II extrinsic factors execute an apoptotic pathway with participation of mitochondria, the significant part of which is shared with apoptosis induced by intrinsic factors. HeLa cells are representatives of the cell type II (Hippe et al., 2008), while the KYM1 cells exhibit conventional apoptosis type I (Schliemann et al., 2011).

The key regulatory role in type I vs type II mechanism is played by the mutual ratio of XIAP vs. procaspase–3 (Aldridge et al., 2011). After the activation of the initiation caspases, apoptosis might proceed by non-mitochondrial pathway (cell type I) provided the ratio of XIAP/procaspase–3 is low. Initiation caspases for cell type I induce direct activation of the effector (executioner) caspases –3 and –7, with lethal outcome for the cell. Reinforcing positive feedback is formed through caspase–3 and –7 activating procaspase–6 and finally the activation of procaspase–8 (Albeck et al., 2008b).

If, however, the ratio between the XIAP and procaspase–3 is high, there is too much XIAP in circulation, the activation of effector caspases–3 and –7 is suppressed by XIAP and the cell proceeds through permeabilization of the external mitochondrial membrane (MOMP). This behavior is defining the cell type II. The example of such behavior is illustrated on Fig. 2. Resulting effector caspases decompose structural proteins of the cell, such as cytokeratines and nuclear lamins, as well as inhibitors of caspase-activated DNAses (iCAD)and poly(ADP-ribose) polymerase (PARP) (Choi et al., 2009). Caspase activated DNAses (CAD) fragment the chromosomal DNA leading to easily observable DNA ladders - hallmarks of late stage apoptosis. The downstream processes after the activation of executioner caspases are not included in the model, though the beginning of this phase is manifested in our model as the cleavage of the PARP protein due to activity of caspase–3.

# Mitochondrial apoptotic pathway

During endocytosis the signaling complex II associates with trans-Golgi vesicules and forms with them multivesicular bodies (MVB). Active caspase–8 activates in MVB ceramid-cathepsin cascade, yielding Cathepsin D cleaving Bid protein to its truncated form tBid (Schütze and Schneider-Brachert, 2009). In the model, these processes are simulated through internalization of the receptor complex I, activation of caspase–8 and truncation of Bid into tBid.

tBid activates Bax. Activated Bax translocates into external mitochondrial membrane where it oligomerizes (often with Bak) and forms permeable pores, through which mitochondria leak critical regulators of apoptosis, such as cytochrome-c and Smac/Diablo, normally localized in mitochondrial intermembrane space.

The MOMP event is usually rapid process, lasting few minutes and it is considered to be the control point, after passing which the cell cannot recover from the apoptotic program (point of no return, "all or nothing" mechanism) (Chipuk et al., 2006). The MOMP transition is clearly seen in the HeLa simulations with current parameters at about 9000 s from beginning (Fig. 2 and Fig.5).

The MOMP transition is attributed to Bcl–2 apoptotic switch (Albeck et al., 2008b) and the transition is triggered when the levels of active Bax/Bak exceed the levels Bcl–2/Bcl-xL, which act as Bax/Bak inhibitors (Korsmeyer et al., 1993). The kinetics of rapid and complete translocation of Smac and cytochrome-c into cytosol through open pores is determined by large concentration gradients, where there is high concentration of these in intermembrane space of mitochondria, while initially practically absent in cytosol (Spencer and Sorger, 2011).

Cytochrome-c dislocated into cytosol forms - together with Apaf–1 and procaspase–9 - apoptosome. Apoptosome cleaves and activates effector procaspases. Available free XIAP associates with catalytic domain of active effector caspases–3 (and caspase–7 not included explicitly in the model), preventing their protease activity and promoting their ubiquitination and thus their degradation. Smac, bound to XIAP, allows to weaken this inhibition, so effector caspases can cleave their substrates and cause cell death (Chipuk et al., 2006).

# Discussion

Single cell experiments using FRET showed, that before MOMP, the activity of effector caspases is negligible, while the initiation caspases are active even before MOMP and the products of their substrates Bid and procaspase–3 accumulate in their cleaved form (Albeck et al., 2008a) (Fig. 5). Caspase–3 is very efficient enzyme: computational simulations supported by experiments indicate, that 400 active molecules are sufficient to cleave $10^6$–$10^7$ molecules of substrate in the course of several hours (Li et al., 2010). The delayed activation of effector caspases is attributed to two mechanisms: competitive XIAP binding to the catalytic site of caspase–3 (which is reversible process) and irreversible ubiquitination and degradation of caspase–3 through E3 ligase XIAP (Albeck et al., 2008b). The time between the cells are exposed to TNF and the MOMP thus represents perhaps the state of “latent death”, where the effector procaspases are continuously activated, but at the same time their active forms are inhibited and the active caspases are effectively degraded through actions of XIAP. This steady-state situation lasts till the release of Smac during MOMP event (Spencer and Sorger, 2011). Then, three possible scenaria might happen: low levels of XIAP allow the effector caspases to cleave the substrates completely, while abundant XIAP inhibits the effector caspases completely. For intermediate levels of XIAP, slow intermediate cleavage, proceeding at suboptimal speed, might appear, leading to genomic instability (Lovric and Hawkins, 2010). The third case can be considered as initiated but unfinished apoptosis. The changes in XIAP levels disturb the normal control of the switch, which controls the activation of effector caspases and disrupts the normal relationships between the activation of effector caspases and the cell death (Spencer and Sorger, 2011). The numerical simulations performed using our model with the HeLa experimental parameters are consistent with the above ideas.

At individual cell level, apoptosis exhibits variability in timing, even amongst the cells of the same population. Single cell experiments show, that some cells die within the 45 minutes after the exposition to cell death signal, while others survive 12 or more hours (Spencer and Sorger, 2011). Individual mitochondria within the same cell differ in their sensitivity to proapoptotic stimuli and MOMP may not fit "all or nothing" picture for all cells (Tait et al., 2010). The MOMP activation thresholds do differ between the cells, reflecting most probably the physiological variations of the Bcl–2 protein family accomplishing the Bcl–2 switch. The computational model presented here allows for variations in individual cell behavior due to variations in internal states and the intensity (Fig. 3) and duration of proapoptotic stimuli. Numerical simulations suggest expected preservation of switching behavior. Lower levels of death stimuli lead to slower Bid truncation and the slower onset of apoptosis. In the case of receptor-mediated apoptosis, MOMP is preceded by a time interval, duration of which is dependent on the dose of stimulus, while the MOMP transition time and post-MOMP course of events are practically dose-independent.

The variability of timing and the probability of apoptosis itself reflects the differences between the cells at genetic and epigenomic levels, different phase of cell cycle, stochastic fluctuations of biochemical reactions and natural variations in protein concentrations. Variations in mRNA levels are reflected in variations of the speed of proteosynthesis. Short-living proteins, or proteins with low number of copies per cell are more vulnerable to such fluctuations, while for more abundant proteins, e.g. part of those controlling apoptosis, the most significant cause is the variability of individual static cell proteome. The explorations of the cell variabilities and their properties encoded in the model will be subject of our future work.

## Conclusion

Predictive modeling of apoptosis, using multifactor, context-sensitive and quantitative data provides the opportunity to distinguish specific responses of individual cell lines and cells to the uniform treatment. Improved and finely tuned models may one day form the base for rational design of individual treatment, tailored toward the medicinal applications. The computational model of TNF signaling pathway seems to provide encouraging results. On the other hand, the amount of specific reliable experimental data needs to grow as well to provide higher level of confidence and improve the predictive power of the models.

# Acknowledgment

The work has been sponsored by grants: VEGA–1/4019/07, OPVaV project ITMS -

26220120024 and APVV–0242–11.

# References

Albeck, J.G., Burke, J.M., Aldridge, B.B., Zhang, M., Lauffenburger, D.A., and Sorger, P.K. (2008a): Quantitative analysis of pathways controlling extrinsic apoptosis in single cells. Mol Cell. **30**, 11–25.

Albeck, J.G., Burke, J.M., Spencer, S.L., Lauffenburger, D.A., and Sorger, P.K. (2008b): Modeling a snap-action, variable-delay switch controlling extrinsic cell death. PLoS Biol **6**, 2831–2852.

Aldridge, B.B., Gaudet, S., Lauffenburger, D.A., and Sorger, P.K. (2011): Lyapunov exponents and phase diagrams reveal multi-factorial control over TRAIL-induced apoptosis. Mol Syst Biol. **7**, 553.

Beck, M., Schmidt, A., Malmstroem, J., Claassen, M., Ori, A., Szymborska, A., Herzog, F., Rinner, O., Ellenberg, J., and Aebersold, R. (2011): The quantitative proteome of a human cell line. Mol Syst Biol. **7**, 7549.

Birbach, A., Gold, P., Binder, B.R., Hofer, E., de Martin, R., and Schmid, J.A. (2002): Signaling molecules of the NF-kappa B pathway shuttle constitutively between cytoplasm and nucleus. J. Biol. Chem. **277**, 10842–10851.

Chen, Z.J. (2012): Ubiquitination in signaling to and activation of IKK. Immunol Rev. **246**, 95–106.

Chipuk, J.E., Bouchier-Hayes, L., and Green, D.R. (2006): Mitochondrial outer membrane permeabilization during apoptosis: the innocent bystander scenario. Cell Death Differ. **13**, 1396–1402.

Choi, E.J., Ahn, W.S., and Bae, S.M. (2009): Equol induces apoptosis through cytochrome c-mediated caspases cascade in human breast cancer MDA-MB-453 cells. Chem Biol Interact. **177**, 7–11.

Feoktistova, M., Geserick, P., Kellert, B., Dimitrova, D.P., Langlais, C., Hupe, M., Cain, K., MacFarlane, M., Häcker, G., and Leverkus, M. (2011): cIAPs block Ripoptosome formation, a RIP1/caspase-8 containing intracellular cell death complex differentially regulated by cFLIP isoforms. Mol. Cell **43**, 449–463.

Gonzalvez, F., and Ashkenazi, A. (2010): New insights into apoptosis signaling by Apo2L/TRAIL. Oncogene **29**, 4752–4765.

Grell, M., Zimmermann, G., Gottfried, E., Chen, C.M., Grünwald, U., Huang, D.C., Wu Lee, Y.H., Dürkop, H., Engelmann, H., Scheurich, P., et al. (1999): Induction of cell death by tumour necrosis factor (TNF) receptor 2, CD40 and CD30: a role for TNF-R1 activation by endogenous membrane-anchored TNF. Embo J **18**, 3034–3043.

Grell, M., Zimmermann, G., Hülser, D., Pfizenmaier, K., and Scheurich, P. (1994): TNF receptors TR60 and TR80 can mediate apoptosis via induction of distinct signal pathways. J. Immunol **153**, 1963–1972.

Heyninck, K., De Valck, D., Vanden Berghe, W., Van Criekinge, W., Contreras, R., Fiers, W., Haegeman, G., and Beyaert, R. (1999): The zinc finger protein A20 inhibits TNF-induced NF-kappaB-dependent gene expression by interfering with an RIP- or TRAF2-mediated transactivation signal and directly binds to a novel NF-kappaB-inhibiting protein ABIN. J Cell Biol. **145**,1471–1482.

Hippe, D., Lytovchenko, O., Schmitz, I., and Lüder, C.G.K. (2008): Fas/CD95-mediated apoptosis of type II cells is blocked by Toxoplasma gondii primarily via interference with the mitochondrial amplification loop. Infect Immun. **76**, 2905–2912.

Hoops, S., Sahle, S., Gauges, R., Lee, C., Pahle, J., Simus, N., Singhal, M., Xu, L., Mendes, P., and Kummer, U. (2006): COPASI--a COmplex PAthway SImulator. Bioinformatics **22**, 3067–3074.

Hucka, M., Finney, A., Sauro, H.M., Bolouri, H., Doyle, J.C., Kitano, H., Arkin, A.P., Bornstein, B.J., Bray, D., Cornish-Bowden, A., et al. (2003): The systems biology markup language (SBML): a medium for representation and exchange of biochemical network models. Bioinformatics **19**, 524–531.

Irmler, M., Thome, M., Hahne, M., Schneider, P., Hofmann, K., Steiner, V., Bodmer, J.L., Schröter, M., Burns, K., Mattmann, C., et al. (1997): Inhibition of death receptor signals by cellular FLIP. Nature **388**, 190–195.

Jupp, O.J., McFarlane, S.M., Anderson, H.M., Littlejohn, A.F., Mohamed, A.A., MacKay, R.H., Vandenabeele, P., and MacEwan, D.J. (2001): Type II tumour necrosis factor-alpha receptor (TNFR2) activates c-Jun N-terminal kinase (JNK) but not mitogen-activated protein kinase (MAPK) or p38 MAPK pathways. Biochem J. **359**, 525–535.

Koh, G., and Lee, D.-Y. (2011): Mathematical modeling and sensitivity analysis of the integrated TNFα-mediated apoptotic pathway for identifying key regulators. Comput Biol Med. **41**, 512–528.

Korsmeyer, S.J., Shutter, J.R., Veis, D.J., Merry, D.E., and Oltvai, Z.N. (1993): Bcl-2/Bax: a rheostat that regulates an anti-oxidant pathway and cell death. Semin Cancer Biol. **4**, 327–332.

Krueger, A., Baumann, S., Krammer, P.H., and Kirchhoff, S. (2001): FLICE-inhibitory proteins: regulators of death receptor-mediated apoptosis. Mol. Cell. Biol. **21,** 8247–8254.

Lavrik, I., Krueger, A., Schmitz, I., Baumann, S., Weyd, H., Krammer, P.H., and Kirchhoff, S. (2003): The active caspase-8 heterotetramer is formed at the CD95 DISC. Cell Death Differ. **10**, 144–145.

Li, C., Donizelli, M., Rodriguez, N., Dharuri, H., Endler, L., Chelliah, V., Li, L., He, E., Henry, A., Stefan, M.I., et al. (2010): BioModels Database: An enhanced, curated and annotated resource for published quantitative kinetic models. BMC Syst Biol. **4**, 92.

Liao, W., Xiao, Q., Tchikov, V., Fujita, K., Yang, W., Wincovitch, S., Garfield, S., Conze, D., El-Deiry, W.S., Schütze, S., et al. (2008): CARP-2 is an endosome-associated ubiquitin ligase for RIP and regulates TNF-induced NF-kappaB activation. Curr Biol. **18**, 641–649.

Lloyd, C.M., Halstead, M.D.B., and Nielsen, P.F. (2004): CellML: its future, present and past. Prog Biophys Mol Biol. **85**, 433–450.

Lovric, M.M., and Hawkins, C.J. (2010): TRAIL treatment provokes mutations in surviving cells. Oncogene **29**, 5048–5060.

Maul, G.G., and Deaven, L. (1977): Quantitative determination of nuclear pore complexes in cycling cells with differing DNA content. J. Cell Biol. **73**, 748–760.

Micheau, O., and Tschopp, J. (2003): Induction of TNF receptor I-mediated apoptosis via two sequential signaling complexes. Cell **114**, 181–190.

Monier, K., Armas, J.C., Etteldorf, S., Ghazal, P., and Sullivan, K.F. (2000): Annexation of the interchromosomal space during viral infection. Nat. Cell Biol. **2**, 661–665.

Nagaraj, N., Wisniewski, J.R., Geiger, T., Cox, J., Kircher, M., Kelso, J., Pääbo, S., and Mann, M. (2011): Deep proteome and transcriptome mapping of a human cancer cell line. Mol. Syst. Biol. **7**, 548.

Neumann, L., Pforr, C., Beaudouin, J., Pappa, A., Fricker, N., Krammer, P.H., Lavrik, I.N., and Eils, R. (2010): Dynamics within the CD95 death-inducing signaling complex decide life and death of cells. Mol. Syst. Biol **6**, 352.

Posakony, J.W., England, J.M., and Attardi, G. (1977): Mitochondrial growth and division during the cell cycle in HeLa cells. J. Cell Biol. **74**, 468–491.

Scheidereit, C. (2006): IkappaB kinase complexes: gateways to NF-kappaB activation and transcription. Oncogene **25**, 6685–6705.

Schliemann, M., Bullinger, E., Borchers, S., Allgöwer, F., Findeisen, R., and Scheurich, P. (2011): Heterogeneity reduces sensitivity of cell death for TNF-stimuli. BMC Syst Biol **5**, 204.

Schütze, S., and Schneider-Brachert, W. (2009): Impact of TNF-R1 and CD95 internalization on apoptotic and antiapoptotic signaling. Results Probl Cell Differ **49**, 63–85.

Spencer, S.L., Gaudet, S., Albeck, J.G., Burke, J.M., and Sorger, P.K. (2009): Non-genetic origins of cell-to-cell variability in TRAIL-induced apoptosis. Nature **459**, 428–432.

Spencer, S.L., and Sorger, P.K. (2011): Measuring and modeling apoptosis in single cells. Cell **144**, 926–939.

Tait, S.W.G., Parsons, M.J., Llambi, F., Bouchier-Hayes, L., Connell, S., Muñoz-Pinedo, C., and Green, D.R. (2010): Resistance to caspase-independent cell death requires persistence of intact mitochondria. Dev. Cell **18**, 802–813.

Tsujimoto, Y., and Shimizu, S. (2005): Another way to die: autophagic programmed cell death. Cell Death Differ. **12** *Suppl 2*, 1528–1534.

Wang, J., Chun, H.J., Wong, W., Spencer, D.M., and Lenardo, M.J. (2001): Caspase-10 is an initiator caspase in death receptor signaling. Proc.Natl.Acad.Sci.U.S.A. **98**, 13884–13888.

Zhao, L., Kroenke, C.D., Song, J., Piwnica-Worms, D., Ackerman, J.J.H., and Neil, J.J. (2008): Intracellular water-specific MR of microbead-adherent cells: the HeLa cell intracellular water exchange lifetime. NMR Biomed **21**, 159–164.

Mitochondria volume - Mammalian tissue culture cell - BNID 100438. Roskams and Rodgers, LabRef;
http://bionumbers.hms.harvard.edu//bionumber.aspx?id=100438&ver=1

# Figure legends

Fig.1. Conceptual scheme of the TNF signaling model.

The individual modules of the reaction network are differentiated through different background. The boundaries between reaction compartments (extracellular space, cytoplasm, nucleus and mitochondria) are marked with lines. Stimulation of the TNF-R1 receptor by TNF-α (TNF) ligand leads to the formation of signaling complexes I (SC1) and II (SC2), which are activating antiapoptotic NFκB and apoptotic signaling pathways. Further antiapoptotic proteins Bcl2_c (cytosolic Bcl–2), Bcl2_m (mitochondrial Bcl–2) and BAR are denoted by gray background. Extracellular space, cytoplasm, nucleus and mitochondria form separate compartments with associated reactions. Mitochondria is marked as an open space, leaking CytC (cytochrome-c) and Smac into surrounding cytosol, due to formation of permeable pores (M#). In the scheme, the symbol $\perp$ denotes an inhibition, while the arrows stand for reactions, activations, transformations and transport between the compartments. To reduce visual clutter, shortened forms of names are used when appropriate (C3 stands for caspase–3, pC9 for procaspase–9, Bcl2_c for cytosolic Bcl2, Bcl2_m for mitochondrial Bcl2 (Bcl-xL)). Full names of the species are specified in the Table 3.

Fig. 2. Simulated time course of the selected molecular concentrations after the application of apoptotic stimulus of 3000 molecules of TNF per HeLa cell. The concentration of antiapoptotic XIAP rises due to activation of antiapoptotic NFκB pathway. After two hours the mitochondrial membrane is permeabilized and apoptotic

proteins (Smac) are released. The drop of XIAP due to interaction of Smac with XIAP is compensated by further synthesis of XIAP (around 10000 seconds from the beginning). After the cytochrome-c is released, apoptosome is formed, and caspase–8 activates very effective caspase–3 (C3). Caspase–3 cleaves the XIAP as well as PARP substrate. The time, when the PARP to cleaved cPARP ratio is 1:1 is sometimes referred as the beginning of apoptosis (here at about 15000 seconds).

Fig. 3. Time dependence of some critical components of apoptotic mechanism - dependence on size of input stimuli. TNF stimulus varies from 120 to 6000 molecules per cell. XIAP concentrations are scaled down by factor 10 (XIAP_norm = [XIAP(t)]/10) to fit on the same figure. Lower levels of input stimuli lead to higher synthesis of XIAP and delayed activity of caspase–3.

Fig. 4. Time course of the molecular concentrations after exposing cells to TNF stimulus. For selected molecular species, the time course is depicted on the figure. On top, the simulations are made with the presented model and the initial amounts corresponding directly to the experimental data taken from proteomics experiment on HeLa cells. On bottom, the same simulation using the KYM1 specific cell line parameters for RIP1, TRADD and TRAF2. Low initial concentrations of these adaptor proteins for HeLa cells cause different activation pattern of FLIP, XIAP and caspase–8. On x-axis the time is specified in seconds, on y-axis amounts of molecular species per reference cell. In order to fit the selected molecules into the same frame, more abundant proteins were scaled down (scaling indicated by _norm suffix). Meaning of the symbols: SC1 and SC2: signaling complexes I and II, respectively. FLIP_norm = [FLIP(t)]/600, XIAP_norm = [XIAP(t)]/125000, C8_norm = [C8(t)]/125000, NFkB_norm =[NFkB(t)]/30

Fig. 5. Why the rise of active caspase–3 concentration is delayed 6000 s after the MOMP event?

Time course of the events after the stimulus 3000 molecules of TNF is applied. Even before MOMP (at 9000 s) we may observe small amount of active caspase–3 (C3_tot comprised of C3:pC6+C3+XIAP:C3) in cytosol. The activity of caspase–3 (C3) is blocked by excessive XIAP before MOMP completely and after the MOMP partially. After MOMP, the amount of XIAP is decreased by two mechanisms: inhibition by (MOMP released) Smac and apoptosome (formed after some delay using MOMP released cytochrome-c) and increased C3 production (C3:XIAP). Finally, the activity of caspase–3 (C3) is notably rised - as evidenced by both absolute count of caspase–3 (C3) and its share on total number of C3_tot. Due to high dynamical range of molecular concentrations, the variables has been scaled to fit the same figure as follows: CytCr_norm = [“cytochrome-C released”(t)]/6, C3_tot_norm = ([C3:pC6(t)]+[C3(t)]+[XIAP:C3(t)])/1100), C3_norm = [C3(t)]/700, XIAP:C3_norm = [XIAP(t)|]/1000, XIAP_tot_norm = ([XIAP(t)]+[Apop:XIAP(t)]+[Smac:XIAP(t)]+[XIAP:C3(t)])/5000, Smac:XIAP_norm=[Smac:XIAP(t)]/1000

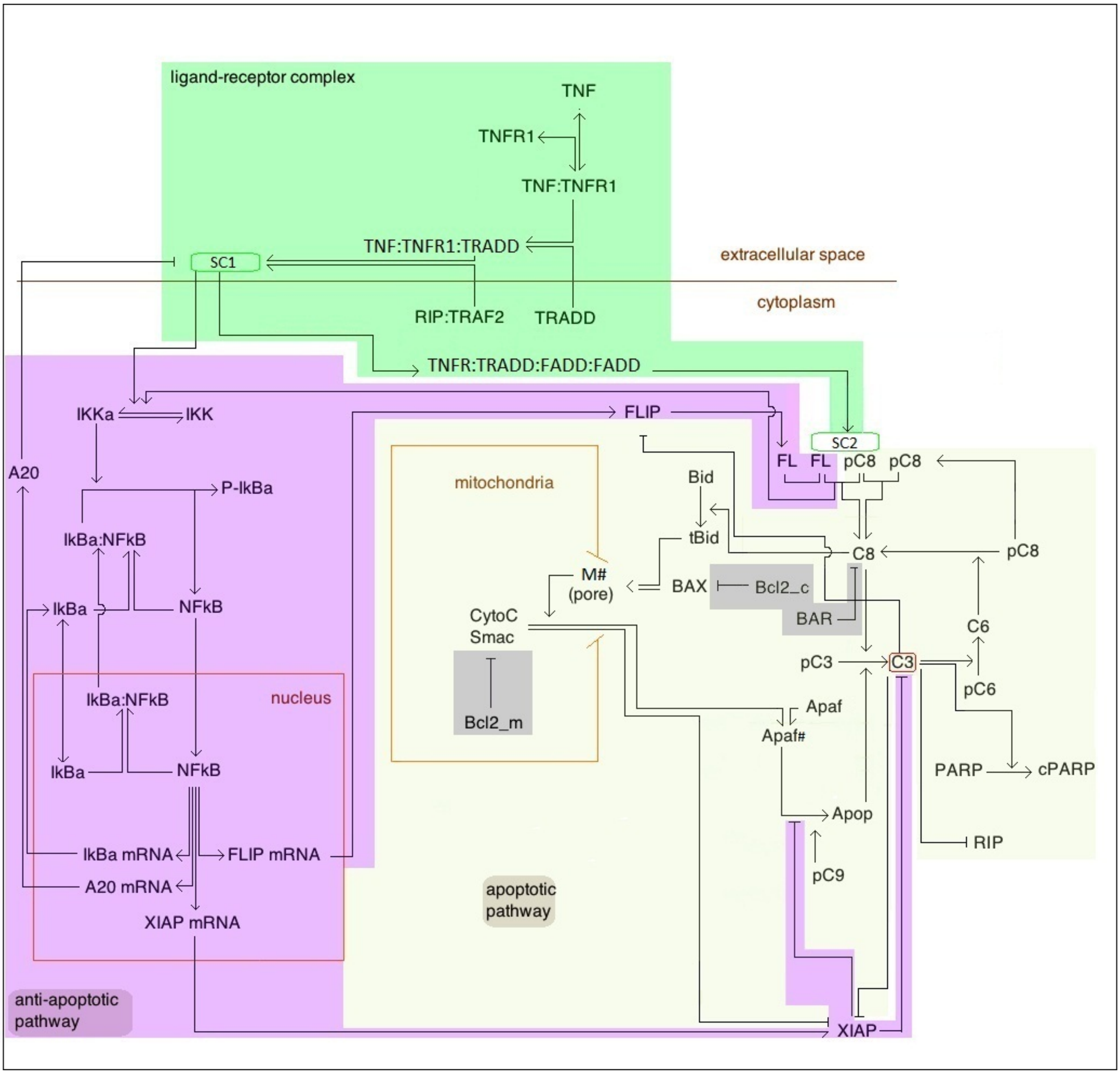

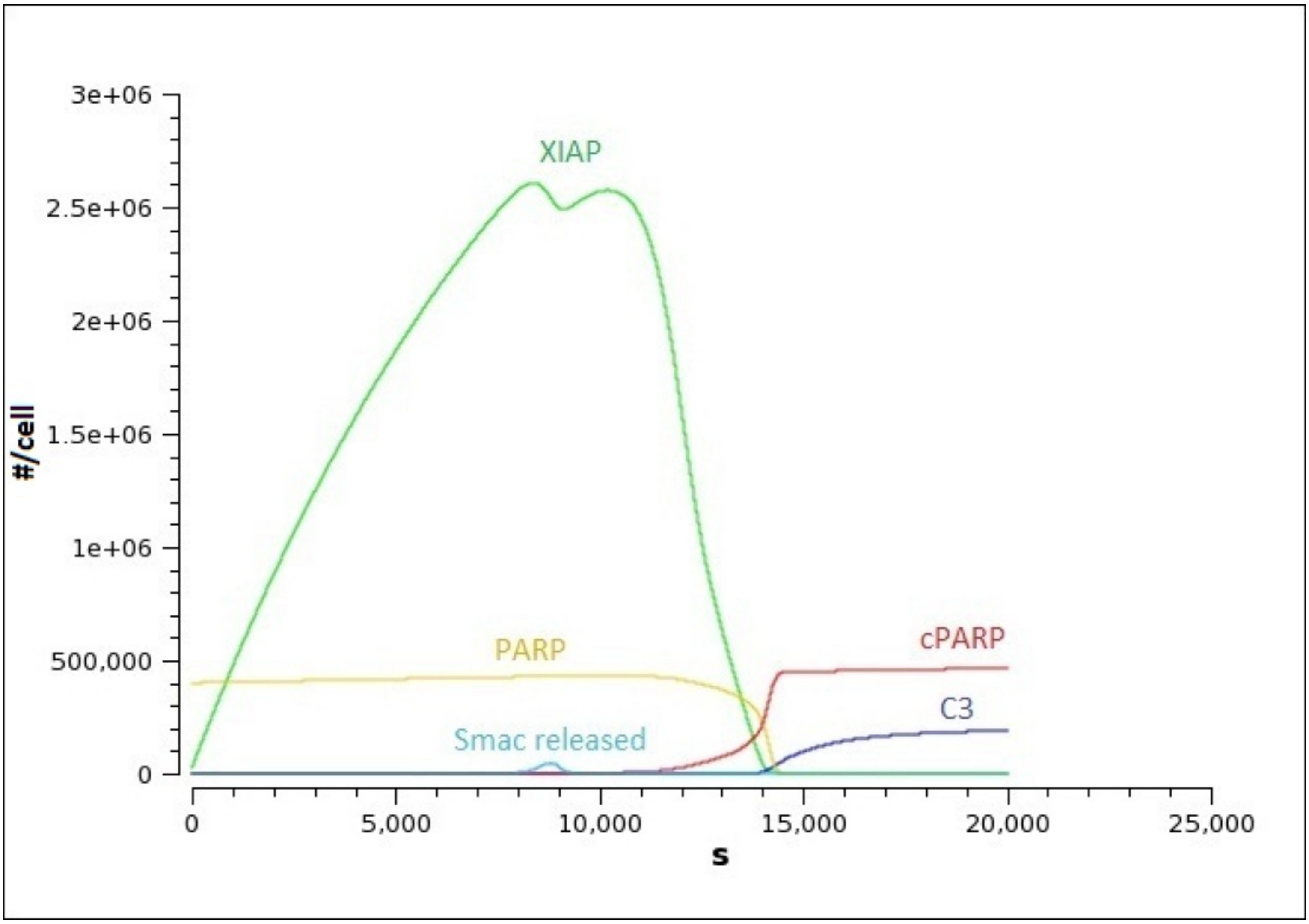

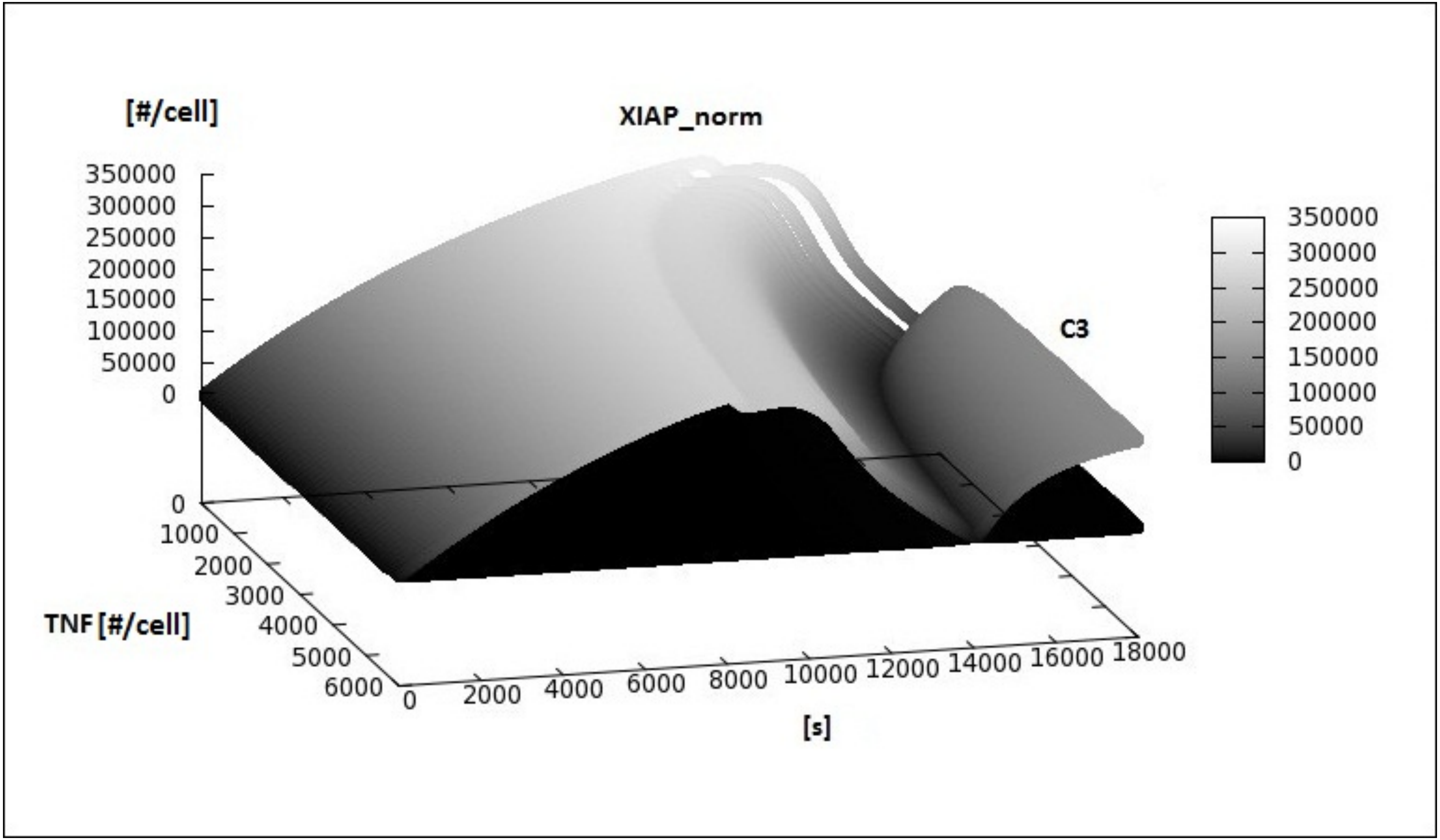

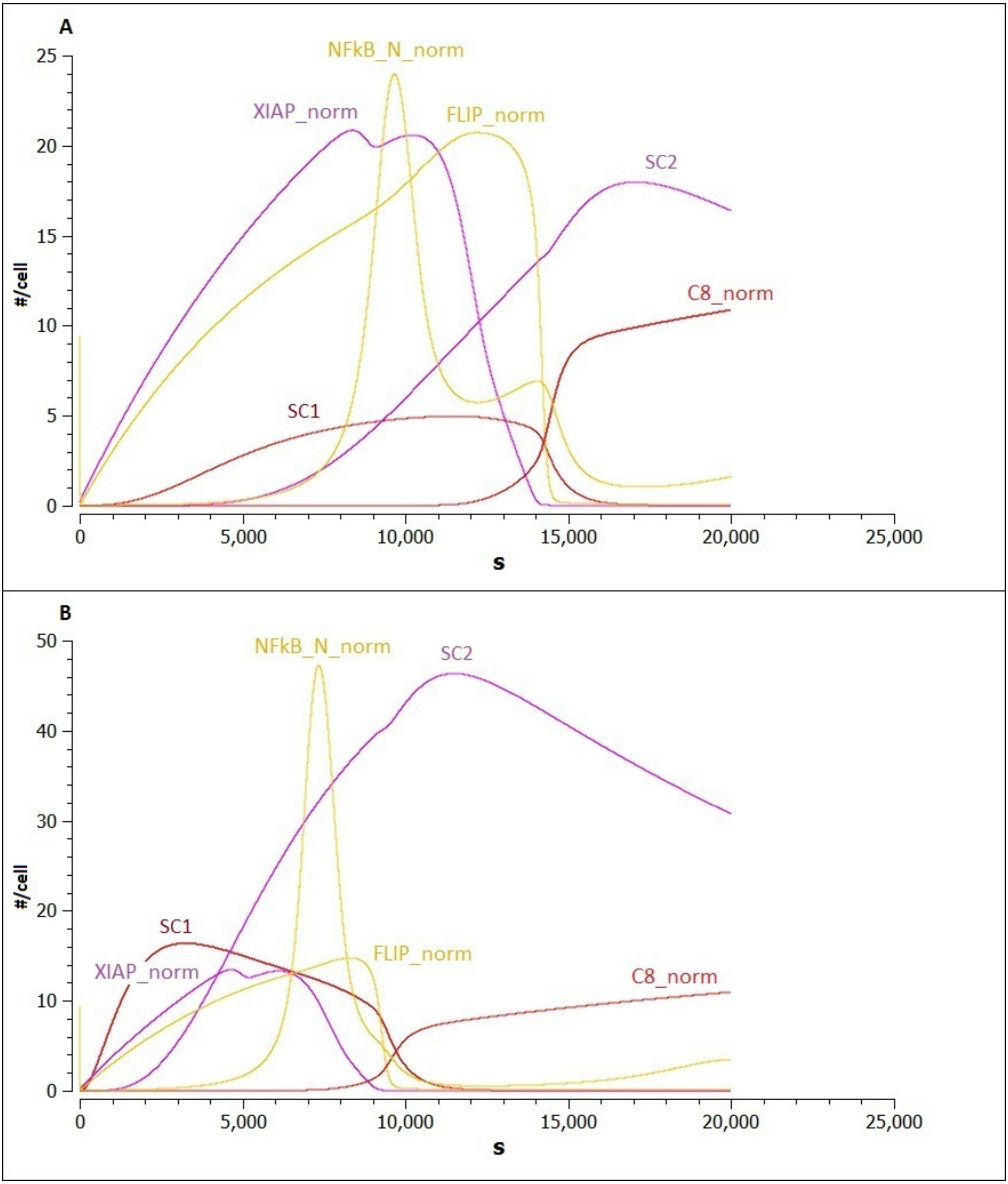

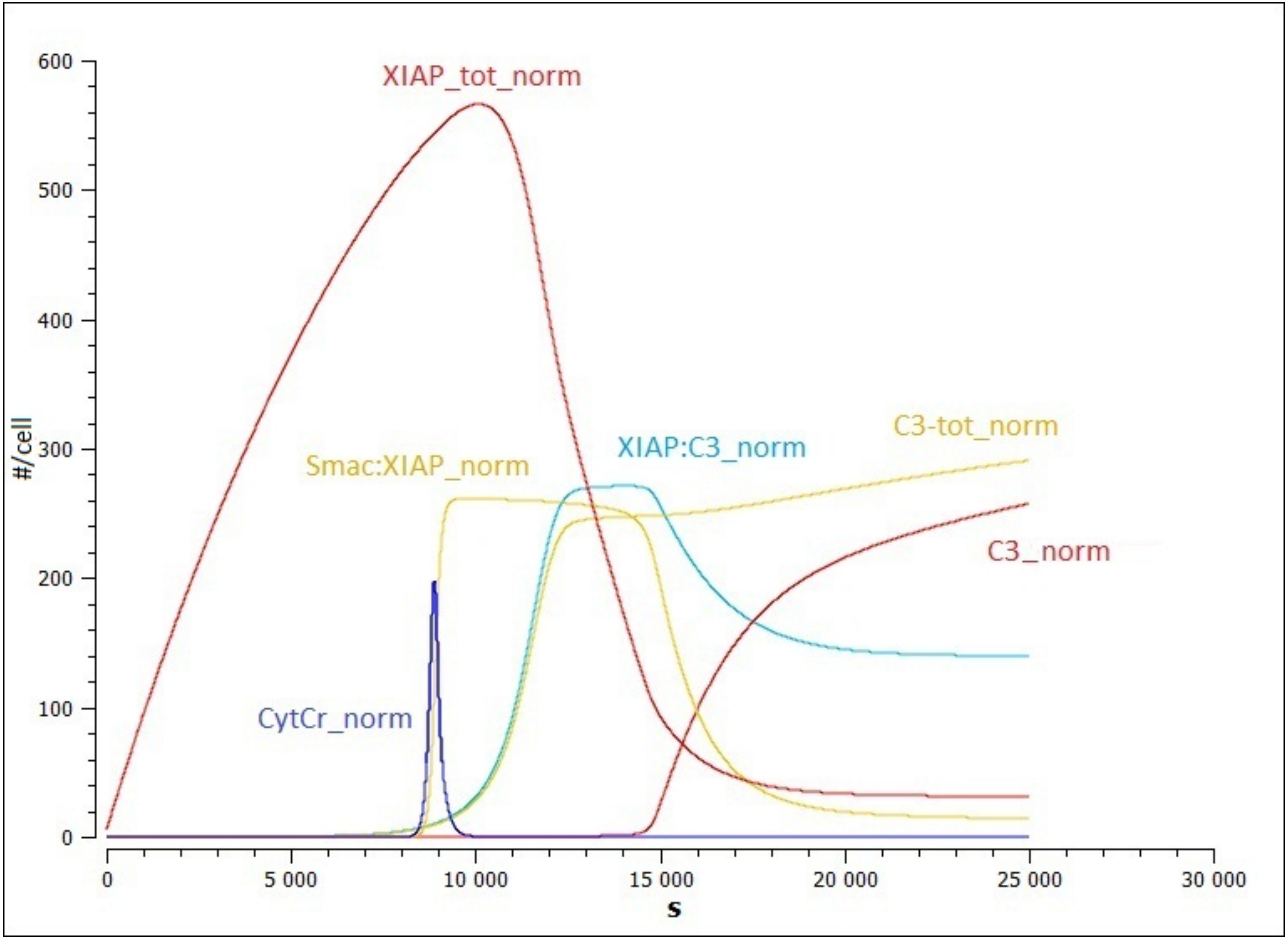

**List of Tables.**

**Table 1. List of main reactions in the model**

| | Reaction[1] | Kinetic parameter | | Compartment volume factor[2] |
|---|---|---|---|---|
| 1 | TNFR → TNFR_E | k1 | | |
| 2 | TNFR_E + TNF_E ⇌ TNF:TNFR_E | k2 | k_2 | |
| 3 | TNF:TNFR_E + TRADD → TNF:TNFR:TRADD | k3 | | |
| 4 | RIP + TRAF2 + TNF:TNFR:TRADD → SC1 | k4 | | |
| 5 | IKK → IKKa; SC1 | k5 | | |
| 6 | IKKa → IKK | k6 | | |
| 7 | SC1 → TRAF2 + TNF:TNFR:TRADD; A20 | k7 | | |
| 8 | NFkB + IkBa → IkBa:NFkB | k8 | | |
| 9 | IkBa:NFkB → NFkB + PIkBa; IKKa | k9 | | |
| 10 | NFkB → NFkB_N | k10 | | |
| 11 | → IkBa_mRNA; NFkB_N | k11 | | VNC |
| 12 | → IkBa; IkBa_mRNA | k12 | | |
| 13 | IkBa ⇌ IkBa_N | k13 | k_13 | |
| 14 | NFkB_N + IkBa_N → IkBa:NFkB_N | k14 | | VNC |
| 15 | IkBa:NFkB_N → IkBa:NFkB | k15 | | |
| 16 | → A20_mRNA; NFkB_N | k16 | | VNC |
| 17 | → A20; A20_mRNA | k17 | | |
| 18 | → XIAP_mRNA; NFkB_N | k18 | | VNC |
| 19 | → XIAP; XIAP_mRNA | k19 | | |
| 20 | → FLIP_mRNA; NFkB_N | k20 | | VNC |
| 21 | → FLIP; FLIP_mRNA | k21 | | |
| 22 | SC1 → Cint1 | k22 | | |
| 23 | Cint1 → RIP + TRAF2 + Cint2 | k23 | | |

[1] The molecular species appearing in reactions are characterized in Table 3. Remaining synthesis and degradation reactions (not listed for brevity) are for all species in the form

Null ⇌ Species

with synthesis rate $k_s$ and degradation rate $k_{deg}$ specified in Table 3.

[2] When listed, VNC (volume of nuclei to cytosol) and VMC (volume of mitochondria to cytosol) correction factors are applied for reaction rates to account for increased speed of reactions in smaller reaction volumes (Albeck et al., 2008).

| | | | | |
|---|---|---|---|---|
| 24 | 2 * FADD + Cint2 → Cint3 | k24 | | |
| 25 | Cint3 → SC2 | k25 | | |
| 26 | SC2 + FLIP → SC2:FL | k26 | | |
| 27 | FLIP + SC2:FL → SC2:FL:FL | k27 | | |
| 28 | SC2 + pC8 → SC2:pC8 | k28 | | |
| 29 | SC2:pC8 + pC8 → SC2:pC8:pC8 | k29 | | |
| 30 | SC2:pC8:pC8 → SC2 + C8 | k30 | | |
| 31 | FLIP + SC2:pC8 → SC2:FL:pC8 | k31 | | |
| 32 | SC2:FL + pC8 → SC2:FL:pC8 | k32 | | |
| 33 | SC2:FL:pC8 → SC2 + C8 | k33 | | |
| 34 | RIP + TRAF2 + SC2:FL:pC8 → SC2:FL:pC8:RIP:TRAF2 | k34 | | |
| 35 | IKK → IKKa;  SC2:FL:pC8:RIP:TRAF2 | k35 | | |
| 36 | BAR + C8 ⇌ BAR:C8 | k36 | k_36 | |
| 37 | pC3 + C8 ⇌ C8:pC3 | k37 | k_37 | |
| 38 | C8:pC3 → C8 + C3 | k38 | | |
| 39 | C3 + pC6 ⇌ C3:pC6 | k39 | k_39 | |
| 40 | C3:pC6 → C3 + C6 | k40 | | |
| 41 | C6 + pC8 ⇌ C6:pC8 | k41 | k_41 | |
| 42 | C6:pC8 → C8 + C6 | k42 | | |
| 43 | XIAP + C3 ⇌ XIAP:C3 | k43 | k_43 | |
| 44 | XIAP:C3 → XIAP | k44 | | |
| 45 | RIP → ;  C3 | k45 | | |
| 46 | FLIP → ;  C3 | k46 | | |
| 47 | PARP → cPARP;  C3 | k47 | | |
| 48 | C8 + Bid ⇌ C8:Bid | k48 | k_48 | |
| 49 | C8:Bid → tBid + C8 | k49 | | |
| 50 | tBid + "cytosolic Bcl-2" ⇌ Bcl2c:tBid | k50 | k_50 | |
| 51 | tBid + Bax ⇌ Bax:tBid | k51 | k_51 | |
| 52 | Bax:tBid → tBid + Bax# | k52 | | |
| 53 | Bax# ⇌ Baxm | k53 | k_53 | |
| 54 | Baxm + Bcl-2 ⇌ Baxm:Bcl2 | k54 | k_54 | VMC |
| 55 | 2 * Baxm ⇌ Bax2 | k55 | k_55 | VMC |
| 56 | 2 * Bax2 ⇌Bax4 | k56 | k_56 | VMC |
| 57 | Bcl-2 + Bax2 ⇌Bax2:Bcl2 | k57 | k_57 | VMC |
| 58 | Bcl-2 + Bax4 ⇌Bax4:Bcl2 | k58 | k_58 | VMC |
| 59 | Bax4 + M ⇌Bax4: | k59 | k_59 | VMC |
| 60 | Bax4:M → M# | k60 | | |
| 61 | M# + CytoC_m ⇌M#:CytoC_m | k61 | k_61 | VMC |
| 62 | M#:CytoC_m → "CytoC released" + M# | k62 | | |
| 63 | M# + Smac_m ⇌M#:Smac_m | k63 | k_63 | VMC |
| 64 | M#:Smac_m → M# + "Smac released" | k64 | | |
| 65 | "CytoC released" ⇌CytoC | k65 | k_65 | |
| 66 | CytoC + Apaf ⇌CytoC:Apaf | k66 | k_66 | |
| 67 | CytoC:Apaf → CytoC + Apaf# | k67 | | |
| 68 | Apaf# + pC9 ⇌Apop | k68 | k_68 | |
| 69 | pC3 + Apop ⇌pC3:Apop | k69 | k_69 | |
| 70 | pC3:Apop → C3 + Apop | k70 | | |
| 71 | "Smac released" ⇌Smac | k71 | k_71 | |
| 72 | XIAP + Apop ⇌Apop:XIAP | k72 | k_72 | |
| 73 | XIAP + Smac ⇌ Smac:XIAP | k73 | k_73 | |
| 74 | M# → M | k74 | | |
| 75 | XIAP →; C3 | k75 | | |

**Table 2. Kinetic rate parameters**

| Name | Value | Units[3] | Source[4] |
|---|---|---|---|
| k1 | 0.001 | $s^{-1}$ | (Schl) |
| k2 | 1.6e-08 | $s^{-1}$*(#/ CC)$^{-1}$ | (Schl) |
| k_2 | 6.60377e-05 | $s^{-1}$ | (Schl) |
| k3 | 7e-09 | $s^{-1}$*(#/ CC)$^{-1}$ | (Schl) |
| k4 | 2.693e-13 | $s^{-1}$*(#/ CC)$^{-2}$ | (Schl) |
| k5 | 0.0001555668 | $s^{-1}$*(#/ CC)$^{-1}$ | (Schl) |
| k6 | 0.1 | $s^{-1}$ | (Schl) |
| k7 | 1e-08 | $s^{-1}$*(#/ CC)$^{-1}$ | (Schl) |
| k8 | 2.076e-06 | $s^{-1}$*(#/ CC)$^{-1}$ | (Schl) |
| k9 | 1.73e-07 | $s^{-1}$*(#/ CC)$^{-1}$ | (Schl) |
| k10 | 0.0125 | $s^{-1}$ | (Schl) |
| k11 | 3.0303e-05 | $s^{-1}$ | (Schl) |
| k12 | 0.0606061 | $s^{-1}$ | (Schl) |
| k13 | 0.0151515 | $s^{-1}$ | (Schl) |
| k_13 | 0.00257576 | $s^{-1}$ | (Schl) |
| k14 | 2.383e-06 | $s^{-1}$*(#/ CC)$^{-1}$ | (Schl) |
| k15 | 0.0151515 | $s^{-1}$ | (Schl) |
| k16 | 3.78788e-05 | $s^{-1}$ | (Schl) |
| k17 | 0.0151515 | $s^{-1}$ | (Schl) |
| k18 | 3.33333e-05 | $s^{-1}$ | (Schl) |
| k19 | 0.0506061 | $s^{-1}$ | (Schl) |
| k20 | 3.33333e-05 | $s^{-1}$ | (Schl) |
| k21 | 0.00687273 | $s^{-1}$ | (Schl) |
| k22 | 0.001135 | $s^{-1}$ | (Schl) |
| k23 | 0.001135 | $s^{-1}$ | (Schl) |
| k24 | 3.27e-14 | $s^{-1}$*(#/ CC)$^{-2}$ | (Schl) |
| k25 | 0.1135 | $s^{-1}$ | (Schl) |
| k26 | 5.19e-07 | $s^{-1}$*(#/ CC)$^{-1}$ | (Schl) |
| k27 | 5.19e-07 | $s^{-1}$*(#/ CC)$^{-1}$ | (Schl) |
| k28 | 5.19e-08 | $s^{-1}$*(#/ CC)$^{-1}$ | (Schl) |
| k29 | 5.19e-08 | $s^{-1}$*(#/ CC)$^{-1}$ | (Schl) |
| k30 | 0.45 | $s^{-1}$ | (Schl) |

[3] #/CC - number of copies of molecule per cell cytosolic volume

[4] Alb – (Albeck et al., 2008), Aldr – (Aldridge et al., 2011), Schl – (Schliemann et al., 2011), Spen – (Spencer and Sorger, 2011)

| | | | |
|---|---|---|---|
| k31 | 5.19e-07 | $s^{-1}*(\#/CC)^{-1}$ | (Schl) |
| k32 | 5.19e-07 | $s^{-1}*(\#/CC)^{-1}$ | (Schl) |
| k33 | 0.3 | $s^{-1}$ | (Schl) |
| k34 | 2.7e-14 | $s^{-1}*(\#/CC)^{-2}$ | (Schl) |
| k35 | 5.2e-08 | $s^{-1}*(\#/CC)^{-1}$ | (Schl) |
| k36 | 8.65e-07 | $s^{-1}*(\#/CC)^{-1}$ | (Schl) |
| k_36 | 0.001 | $s^{-1}$ | (Schl) |
| k37 | 1e-07 | $s^{-1}*(\#/CC)^{-1}$ | (Alb) |
| k_37 | 0.001 | $s^{-1}$ | (Alb) |
| k38 | 1 | $s^{-1}$ | (Alb) |
| k39 | 1e-07 | $s^{-1}*(\#/CC)^{-1}$ | (Spen) |
| k_39 | 0.001 | $s^{-1}$ | (Alb) |
| k40 | 1 | $s^{-1}$ | (Alb) |
| k41 | 1e-07 | $s^{-1}*(\#/CC)^{-1}$ | (Spen) |
| k_41 | 0.001 | $s^{-1}$ | (Alb) |
| k42 | 1 | $s^{-1}$ | (Alb) |
| k43 | 1.038e-06 | $s^{-1}*(\#/CC)^{-1}$ | (Alb) |
| k_43 | 0.001 | $s^{-1}$ | (Alb) |
| k44 | 5e-05 | $s^{-1}$ | (Alb) |
| k45 | 2.59e-07 | $s^{-1}*(\#/CC)^{-1}$ | (Alb) |
| k46 | 2.59e-07 | $s^{-1}*(\#/CC)^{-1}$ | (Alb) |
| k47 | 3.11e-07 | $s^{-1}*(\#/CC)^{-1}$ | (Alb) |
| k48 | 1e-07 | $s^{-1}*(\#/CC)^{-1}$ | (Alb) |
| k_48 | 0.001 | $s^{-1}$ | (Alb) |
| k49 | 1 | $s^{-1}$ | (Alb) |
| k50 | 1e-06 | $s^{-1}*(\#/CC)^{-1}$ | (Alb) |
| k_50 | 0.001 | $s^{-1}$ | (Alb) |
| k51 | 1e-07 | $s^{-1}*(\#/CC)^{-1}$ | (Alb) |
| k_51 | 0.001 | $s^{-1}$ | (Alb) |
| k52 | 1 | $s^{-1}$ | (Alb) |
| k53 | 0.01 | $s^{-1}$ | (Alb) |
| k_53 | 1 | $s^{-1}$ | (Spen) |
| k54 | 1e-06 | $s^{-1}*(\#/CC)^{-1}$ | (Alb) |
| k_54 | 0.001 | $s^{-1}$ | (Alb) |
| k55 | 1e-06 | $s^{-1}*(\#/CC)^{-1}$ | (Alb) |
| k_55 | 0.001 | $s^{-1}$ | (Alb) |
| k56 | 1e-06 | $s^{-1}*(\#/CC)^{-1}$ | (Alb) |
| k_56 | 0.001 | $s^{-1}$ | (Alb) |
| k57 | 1e-06 | $s^{-1}*(\#/CC)^{-1}$ | (Alb) |
| k_57 | 0.001 | $s^{-1}$ | (Alb) |
| k58 | 1e-06 | $s^{-1}$*(#/ CC)$^{-1}$ | (Alb) |
| k_58 | 0.001 | $s^{-1}$ | (Alb) |
| k59 | 1e-06 | $s^{-1}$*(#/ CC)$^{-1}$ | (Alb) |
| k_59 | 0.001 | $s^{-1}$ | (Alb) |
| k60 | 1 | $s^{-1}$ | (Alb) |
| k61 | 2e-06 | $s^{-1}$*(#/ CC)$^{-1}$ | (Alb) |
| k_61 | 0.001 | $s^{-1}$ | (Alb) |
| k62 | 10 | $s^{-1}$ | (Alb) |
| k63 | 2e-06 | $s^{-1}$*(#/ CC)$^{-1}$ | (Alb) |
| k_63 | 0.001 | $s^{-1}$ | (Alb) |
| k64 | 10 | $s^{-1}$ | (Alb) |
| k65 | 1 | $s^{-1}$ | (Spen) |
| k_65 | 0.01 | $s^{-1}$ | (Alb) |
| k66 | 5e-07 | $s^{-1}$*(#/ CC)$^{-1}$ | (Alb) |
| k_66 | 0.001 | $s^{-1}$ | (Alb) |
| k67 | 1 | $s^{-1}$ | (Alb) |
| k68 | 5e-08 | $s^{-1}$*(#/ CC)$^{-1}$ | (Alb) |
| k_68 | 0.001 | $s^{-1}$ | (Alb) |
| k69 | 5e-09 | $s^{-1}$*(#/ CC)$^{-1}$ | (Alb) |
| k_69 | 0.001 | $s^{-1}$ | (Alb) |
| k70 | 1 | $s^{-1}$ | (Alb) |
| k71 | 1 | $s^{-1}$ | (Spen) |
| k_71 | 0.01 | $s^{-1}$ | (Alb) |
| k72 | 2e-06 | $s^{-1}$*(#/ CC)$^{-1}$ | (Alb) |
| k_72 | 0.001 | $s^{-1}$ | (Alb) |
| k73 | 7e-06 | $s^{-1}$*(#/ CC)$^{-1}$ | (Alb) |
| k_73 | 0.001 | $s^{-1}$ | (Alb) |
| k74 | 0.0001 | $s^{-1}$ | (Aldr) |
| k75 | 3.114 | $s^{-1}$*(#/ CC)$^{-1}$ | (Schl) |
| VNC | 0,25 | | |
| VMC | 0,1 | | |

**Table 3. Initial conditions, synthesis and degradation rates**

| Symbol | #/CC source: MS | $k_s$ (#* $s^{-1}$) | $k_{deg}$ ($s^{-1}$) | Source of reaction rates[5] | Notes | Com part ment[6] |
|---|---|---|---|---|---|---|
| TNFR | 168 | 0.168 | | (Schl) | Newly synthetized TNF receptor | CC |
| RIP | 9500 | 12 | 0.0001 | (Schl) | *Receptor-interacting protein* (RIPK1) | CC |
| TRADD | 50000 | 17.7 | 0.0001 | (Schl) | *TNFR1-associated DEATH domain protein* (TRADD) | CC |
| TRAF2 | 13000 | 19.9 | 0.0001 | (Schl) | *TNF receptor-associated factor 2* (TRAF2) | CC |
| FADD | 18000 0 | 18.6 | 0.0001 | (Schl) | *FAS-associated death domain protein* (FADD) | CC |
| TNF:TNFR:TRADD | 0 | | 0.02352 | (Schl) | TNF~TNFR~TRADD complex | CC |
| SC1 | 0 | | 5.6e-05 | (Schl) | TNF-R1-complex containing RIP and TRAF2 (signaling complex I) | CC |
| Cint1 | 0 | | | | internalized signaling complex I, state 1 | CC |
| Cint2 | 0 | | | | internalized signaling complex I, state 2 | CC |
| Cint3 | 0 | | | | internalized signaling complex I, state 3 | CC |
| SC2 | 0 | | 5.6e-05 | (Schl) | TNF-R1-complex containing FADD (signaling complex II) | CC |
| FLIP | 0 | 1.354348 | 0.0001 | (Schl) | *FLICE-inhibitory protein* (CFLAR) | CC |
| SC2:FL | 0 | | 5.6e-05 | (Schl) | SC2~FLIP complex | CC |
| SC2:pC8 | 0 | | 5.6e-05 | (Schl) | SC2~procaspase-8 complex | CC |
| SC2:FL:FL | 0 | | 5.6e-05 | (Schl) | SC2~FLIP~FLIP complex | CC |
| SC2:pC8:pC8 | 0 | | 5.6e-05 | (Schl) | SC2~procaspase-8~procaspase-8 complex | CC |
| SC2:FL:pC8 | 0 | | 5.6e-05 | (Schl) | SC2~FLIP~procaspase-8 complex | CC |

[5] Alb – (Albeck et al., 2008), Aldr – (Aldridge et al., 2011), Schl – (Schliemann et al., 2011), Spen – (Spencer and Sorger, 2011)

[6] compartments: CC – cytosol, MC – mitochondria, NC – nucleus, EC- extracelular space. The size of EC was set to equal CC.

| | | | | | | |
|---|---|---|---|---|---|---|
| SC2:FL:pC8:RIP:TRAF2 | 0 | | 5.6e-05 | (Schl) | SC2~p43-FLIP complex with RIP and TRAF2 | CC |
| IKK | 3500 | 38.541 | 0.0001 | (Schl) | *Inhibitor-kB kinase*, inactive form [IKK1 (CHUK), IKK2 (IKBKB), NEMO (IKBKG)] | CC |
| IKKa | 0 | | 0.0001 | (Schl) | active form IKK | CC |
| A20 | 1600 | 5.78 | 0.0001 | (Schl) | *Tumor necrosis factor, alpha-induced protein 3* (TNFAIP3) | CC |
| NFkB | 75 | 0.96352 | 0.0001 | (Schl) | *Nuclear factor κB* [p50 (NFKB1), p65 (RELA)] | CC |
| IkBa | 67060 | | 0.00154022 | (Schl) | *Inhibitor-κBα* (NFKBIA) | CC |
| IkBa:NFkB | 5840 | | 0.0001 | (Schl) | IκBα ~NF-κB complex | CC |
| PIkBa | 0 | | 0.0115517 | (Schl) | Phosphorylated form of IκBα | CC |
| BAR | 0 | 1 | 5.78704e-05 | (Schl) | *Bifunctional apoptosis regulator* (BFAR) | CC |
| XIAP | 27000 | 465 | 0.0001 | (Schl) | *X-linked inhibitor of apoptosis protein* (XIAP) | CC |
| pC8 | 70000 | 118.9531682 | 6.17284e-05 | (Schl) | procaspase-8 (CASP8) | CC |
| pC3 | 64000 | 29.738262 | 6.17284e-05 | (Schl) | procaspase-3 (CASP3) | CC |
| pC6 | 12000 | 2.379 | 6.17284e-05 | (Schl) | procaspase-6 (CASP6) | CC |
| C8 | 0 | | 5.78704e-05 | (Schl) | caspase-8 | CC |
| C3 | 0 | | 5.78704e-05 | (Schl) | caspase-3 | CC |
| C6 | 0 | | 5.78704e-05 | (Schl) | caspase-6 | CC |
| BAR:C8 | 0 | | 5.78704e-05 | (Schl) | BAR~caspase-8 complex | CC |
| XIAP:C3 | 0 | | 5.78704e-05 | (Schl) | XIAP~caspase-3 complex | CC |
| PARP | 400000 | 5.808 | 5.78704e-06 | (Schl) | *Poly [ADP-ribose] polymerase1* (PARP1) 320000#, lumped PARP species | CC |
| cPARP | 0 | | 5.78704e-06 | (Schl) | Cleaved PARP | CC |
| C6:pC8 | 0 | | 2.9e-06 | (Spen) | Caspase-6~procaspase-8 complex | CC |
| C8:pC3 | 0 | | 2.9e-06 | (Spen) | Caspase-8~procaspase-3 complex | CC |
| Bid | 440000 | 0.02610 | 2.9e-06 | (Spen) | *BH3 interacting domain death agonist* (BID) | CC |
| C8:Bid | 0 | | 2.9e-06 | (Spen) | Caspase-8~Bid komplex | CC |
| tBid | 0 | | 2.9e-06 | (Spen) | truncated Bid | CC |
| Apop | 0 | | 2.9e-06 | (Spen) | apoptosome | CC |
| pC3:Apop | 0 | | 2.9e-06 | (Spen) | procaspase-3~apoptosome complex | CC |
| C3:pC6 | 0 | | 2.9e-06 | (Spen) | Caspase-3~procaspase-6 complex | CC |
| Apop:XIAP | 0 | | 2.9e-06 | (Spen) | Apoptosome~XIAP complex | CC |
| Smac | 0 | | 2.9e-06 | (Spen) | cytosolic Smac | CC |
| Smac:XIAP | 0 | | 2.9e-06 | (Spen) | Smac~XIAP complex | CC |
| "cytosolic Bcl-2" | 25000 | 0.00870 | 0.0001 | (Spen) | *B-cell CLL/lymphoma 2* (BCL2) | CC |
| Bax | 500000 | 0.03480 | 2.9e-06 | (Spen) | *BCL2-associated X protein* (BAX) | CC |
| Bax:tBid | 0 | | 2.9e-06 | (Spen) | Bax~tBid complex | CC |
| Bax# | 0 | | 2.9e-06 | (Spen) | active form Bax | CC |
| CytoC | 0 | | 2.9e-06 | (Spen) | cytochromee c | CC |
| Apaf | 3300 | 0.04350 | 2.9e-06 | (Spen) | *Apoptotic peptidase activating factor1* (APAF1) | CC |
| CytoC:Apaf | 0 | | 2.9e-06 | (Spen) | CytoC~Apaf complex | CC |
| Apaf# | 0 | | 2.9e-06 | (Spen) | active form Apaf | CC |
| pC9 | 6500 | 0.04350 | 2.9e-06 | (Spen) | Procaspase-9 (CASP9) | CC |
| Bcl2c:tBid | 0 | | 2.9e-06 | (Spen) | Bcl2~tBid complex | CC |
| TNFR_E | 2850 | | 5.6e-05 | (Schl) | TNFR1 dimer (TNFRSF1A), (TNFRSF1B), both are missing in MS data | EC |
| TNF_E | 3000 | | 2.9e-06 | (Schl) | stimulus *Tumor Necrosis Factor* (TNF) | EC |
| TNF:TNFR_E | 0 | | 5.6e-05 | (Schl) | ligand~receptor complex | EC |
| NFkB_N | 251 | | 0.0001 | (Schl) | nuclear NF-κB | NC |
| IkBa_N | 16766 | | 0.0001 | (Schl) | nuclear IκBα | NC |
| IkBa:NFkB_N | 334 | | 0.0001 | (Schl) | IκBα ~NF-κB complex in nucleus | NC |
| A20_mRNA | 34 | | 0.000470498 | (Schl) | A20-encoding mRNA | NC |
| IkBa_mRNA | 32 | | 0.000394201 | (Schl) | IκBα-encoding mRNA | NC |
| XIAP_mRNA | 132 | | 0.000104931 | (Schl) | XIAP-encoding mRNA | NC |
| FLIP_mRNA | 84 | | 0.000165744 | (Schl) | FLIP-encoding mRNA | NC |
| Bax2 | 0 | | 2.9e-06 | (Spen) | Bax dimer | MC |
| Bax2:Bcl2 | 0 | | 2.9e-06 | (Spen) | Bax2~Bcl2 complex | MC |
| Bax4 | 0 | | 2.9e-06 | (Spen) | Bax tetramer | MC |
| Bax4:Bcl2 | 0 | | 2.9e-06 | (Spen) | Bax4~Bcl2 complex | MC |
| Bax4:M | 0 | | 2.9e-06 | (Spen) | Bax4~pore komplex | MC |
| Baxm | 0 | | 2.9e-06 | (Spen) | Bax in mitochondria | MC |
| Baxm:Bcl2 | 0 | | 2.9e-06 | (Spen) | Baxm~Bcl2 complex | MC |
| Bcl-2 | 60000 | 0.00870 | 2.9e-06 | (Spen) | *BCL2-like 1* (BCL2L1) | MC |
| "CytoC released" | 0 | | 2.9e-06 | (Spen) | cytochrome c released but remaining in the MC | MC |
| CytoC_m | 160000 | 0.21750 | 2.9e-06 | (Spen) | cytochrome c (CYCS) | MC |
| M | 720000 | 0.21750 | 2.9e-06 | (Spen) | *Translocator protein* pore (TSPO) | MC |
| M# | 0 | | | | Activated pore | MC |
| M#:Smac_m | 0 | | 2.9e-06 | (Spen) | pore~Smac_m complex | MC |
| "Smac released" | 0 | | 2.9e-06 | (Spen) | Smac released but remaining in the MC | MC |
| M#:CytoC_m | 0 | | 2.9e-06 | (Spen) | M#~CytoC_m complex | MC |
| Smac_m | 270000 | 0.04350 | 2.9e-06 | (Spen) | *IAP-binding mitochondrial protein* (DIABLO) | MC |

**Table 4. Initial concentration of selected species in three different models**

| Molecule | Mass Spectrometry, HeLa, #/cell | Schliemann, KYM1, #/cell | Spencer, HeLa, #/cell |
|---|---|---|---|
| RIP | 9500 | 121985 | |
| TRADD | 50000 | 176714 | |
| TRAF2 | 13000 | 199068 | |
| FLIP | 0 | 19299 | 2000 |
| IKK | 3500 | 385417 | |
| A20 | 1600 | 62891 | |
| BAR | 0 | 173371 | 1000 |
| XIAP | 27000 | 4717570 | 100000 |
| pC8 | 70000 | 1927090 | 10000 |
| pC3 | 64000 | 481771 | 10000 |
| pC6 | 12000 | 38542 | 10000 |
| IkBa (total) | 90000 | 7253 | |
| NFkB (total) | 6500 | 5621 | |
| BID | 440000 | | 60000 |
| BAX | 500000 | | 80000 |
| CytoC_m | 160000 | | 500000 |
| Smac_m | 270000 | | 100000 |
| Apaf | 3300 | | 100000 |
| pC9 | 6500 | | 100000 |